\magnification=\magstep1
\baselineskip 16pt 

\font\hdl=cmfib8 scaled \magstep2
\font\bet=cmfib8 scaled \magstep1

\newcount\refnumber
\refnumber=0
\immediate
\def\ref#1{\global\advance\refnumber by 1}

\newcount\equnumbera
\def\eqna{\eqno(\the\equnumbera)}

\immediate
\def\equorder#1{\global\advance\equnumbera by 1}

\def\fqpt{f(Q,P,t)}
\def\xvi{{\hat x}_i}
\def\pvi{{\hat p}_i}
\def\om{\omega }
\def\hbo{\hbar \om}
\def\zerofri{\gamma(0)}
\def\h#1{{\hat #1}}
\def\drft{{\tilde \chi}^{\prime \prime }}
\def\drf{\chi^{\prime \prime }}
\def\rft{{\tilde \chi}}

\def\chiosc{\chi_{\rm osc}(\om)}
\def\bra{\big\langle}
\def\ket{\big\rangle}
\def\fmb{ \langle \hat F \rangle }
\def\ham#1{\hat H(\xvi,\pvi,#1)}

\def\ffield#1{ \hat F(\xvi,\pvi,#1)}
\def\equiop{\hat{\rho}_{\rm qs}}

\def\dtwohqzero{\left<{\partial^2\hat H\over \partial Q^2}(Q_0)
    \right>^{\rm qs}_{Q_0,T_0}} 
\def\fext{f_{\rm ext}}
\def\respc{\chi_{\rm coll}(\om)}
\def\dpr{{\prime \prime}}
\def\corrheat{_0\psi^{\dpr }}
\def\self{{\it\Sigma}}
\def\selfre{\Sigma^\prime}


\def\traneq{$${\partial  \over \partial t} \fqpt = \Biggl[
    -{\partial\over\partial Q}{P\over M} + {\partial\over\partial P} 
     \left({d V(Q)\over d Q}\right)
    +{\partial\over\partial P}{P\over M}\gamma +
     D_{pp}{\partial^2\over\partial P\partial P} \Biggr] \fqpt
    \eqna $$} 
\def\propagdens{$$ f(Q,P,t) = \int {\rm d}Q_0 {\rm d}P_0 \; \; 
    K(Q,P,t;Q_0,P_0,t_0) \; \; f(Q_0,P_0,t_0) \eqna $$}

\def\decolldiff{$$-{d\over dt} E_{\rm coll}  \equiv
    -{d\over dt}\left({M(\om_{1})\over 2}{\dot q}^{2}
    +{C(\om_{1})\over 2}q^{2}\right) = \gamma(\om_{1}){\dot q}^{2}
    \equiv  T {d\over dt}S 
    \eqna $$ }
\def\dspspecdensgam{$$ \varrho_k = {\Gamma(\om) \over
    \left(\hbar \om -e_k -\selfre(\om)\right)^2 + 
    \left({\Gamma(\om)\over 2} \right)^2}
\qquad \Gamma=
    {1\over \Gamma_0}\;{(\hbo - \mu)^2 + \pi^2 T^2 \over 
    1 + \left[(\hbo - \mu)^2 + \pi^2 T^2 \right] /c^2}
    \eqna $$}
\def\hamilapp{$$\hat H(Q(t))=\hat H(Q_0)  + 
     (Q(t)-Q_0)\h{F} +
     {1\over 2}(Q(t)-Q_0)^2 \dtwohqzero  \eqna $$} 
\def\eqofmotim{$$ 0 = {d \over dt} E_{\rm tot} = {\dot Q}
   \left<{\partial \ham{Q} \over \partial Q}\right>_{t} \equiv 
    {\dot Q} \left< \ffield{Q} \right>_{t}
    \eqna $$} 
\def\defrest{$$ \rft (t-s)= \Theta(t-s) {i\over \hbar} 
             {\rm tr}\,\left(\hat{\rho}_{\rm qs}(Q_{0},T_{0})
             [\hat F^{I}(t),\hat F^{I}(s)] \right) 
     \equiv 2 i \Theta(t-s) \drft(t-s)
             \eqna $$}
\def\tranosc{$$ \left(\respc\right) ^{-1}\; \delta \fmb_\om  \simeq
    \left(\chiosc \right) ^{-1}\; \delta \fmb_\om  \equiv
    \left(-M\om^2 -\gamma i \om + C\right)\; \delta \fmb_\om 
    = -  \fext(\om)    \eqna $$ }
\def\zefridisflu{$$ \zerofri=
     {{\partial \chi^{\dpr} (\om)} \over{\partial \om}}
     \Bigr\arrowvert_{\om = 0}=\Phi^{\prime\prime}(\om=0)=
     {\psi^{\dpr}(\om=0) \over 2T} 
     \eqna $$ }

\def\corrsingreg{$$ \psi^{\dpr }(\om)= \psi^{0}
    2\pi \delta(\om) \; + \;_R\psi^{\dpr }(\om)
    \qquad {\rm with} \qquad \psi^{0}= T\Bigl( \chi^{\rm T} - \chi(0) \Bigr)
    \eqna$$}
\def\sepchiadtis{$${1\over T}\psi^{0} = \chi^{\rm T} - \chi(0)  =
    \Bigl( \chi^{\rm T} - \chi^{\rm ad} \Bigr)-
    \Bigl( \chi^{\rm ad} - \chi(0) \Bigr) \qquad \longrightarrow \qquad
   {1\over T}\psi^{0}  = \chi^{\rm T} - \chi^{\rm ad}   \eqna $$}
\def\heatplor{$$  \corrheat(\om)= \psi^0 2\pi \delta(\om) \qquad
    \Longrightarrow \qquad \corrheat(\om) = \psi^0 {{\hbar \Gamma_T}\over 
    {\hbar^2\om^2 +\Gamma_T^2 /4}}    \eqna $$}
\def\zerfriheat{$$ {_0\zerofri}= {4 \hbar \over \Gamma_T}{\psi^{0}\over
    2T} = {2 \hbar \over \Gamma_T}\Bigl( \chi^{\rm T} - \chi(0) \Bigr)
    \eqna $$ }
\def\avernlos{$$ \overline {\Delta
   E_{int}}=-\int_{-T/2}^{T/2}dt\int_{-\infty}^{\infty} 
   ds\, \dot q(t)\tilde \chi(t-s)q(s) =
   \pi q_0^2 \chi^{\prime \prime}(\Omega)
    \eqna$$ }
\def\avenlosfric{$$ \overline {{d E_{int}\over dt}}
    ={{\pi}q^2_0\chi^{\prime\prime}(\Omega)\over (2\pi/\Omega)}
    ={{\overline v}^2\chi^{\prime\prime}(\Omega)\over\Omega}
   =\gamma (\Omega)\,{\overline v}^2  \eqna $$ }
\def\fricenlo{$$\gamma(\Omega) \equiv {\chi^{\prime\prime}(\Omega)
   \over \Omega} 
    \equiv \Phi^{\prime\prime}(\Omega) \eqna $$ }


\centerline{{\hdl Warm nuclei: The transition from }}
\medskip
\centerline{{\hdl independent particle motion to collisional dominance}
\footnote{$^1$}
{Supported in part by the Deutsche Forschungsgemeinschaft}
}

\medskip
\centerline{by}
\medskip
\centerline{{\bet Helmut Hofmann}
\footnote{$^2$}
{e-mail: hhofmann@physik.tu-muenchen.de}
}
\centerline{Physik Department, TUM, D-85747 Garching}
\medskip
\centerline{{\bet Fedor A. Ivanyuk and Alexander G. Magner}}
\centerline{Institute for Nuclear Research, 252028, Kiev-28}
\centerline{Physik Department, TUM, D-85747 Garching}

\vskip 1 cm

\par{\parindent 15pt
\narrower
We study large scale collective dynamics of isoscalar type and examine
the influence of interactions residual to independent particle motion.
It is argued that for excitations which commonly are present in
experimental situations such interactions must not be neglected. They
even help to justify better the assumption of locality, both in time as
well as in phase space, which is necessary not only for such classic
approaches to collective motion as the "cranking model" but also for
the more general picture of a transport process. With respect to
dissipation, our results are contrasted with those of wall friction.
\par}

\bigskip
\centerline{PACS 21.60.Ev, 21.60.Cs, 24.10.Pa, 24.75+i}

\vskip 1 cm
\leftline{{\bf 1 Introduction}\hfill }
\medskip
\ref{kram}
\edef\nkram{\the\refnumber}

After the discovery of the shell model it has become customary
to base the description of collective motion on the picture of single
particles moving independently within a deformed mean field. This
approach was introduced in the early 50'ties by A. Bohr and B. Mottelson to
portray low energetic collective excitations, and to the present day
there can be little doubt that this approximation is adequate for that
regime. It is somewhat astonishing, however, that this picture still is
vindicated by many groups even for situations where the nucleons are
heated up to considerable amount, say to temperatures of a few MeV.
After all, in the very early days of nuclear physics collective motion
of large scale was considered to be governed by dissipative processes,
which in turn imply the presence of fluctuating forces. Such a picture
may be condensed into the one equation, which was suggested by Kramers
[\nkram] already in 1940 to describe nuclear fission. It reads
\equorder{traneq}
\traneq
\edef\ntraneq{\the\equnumbera}
and has the structure typical of a Fokker-Planck equation. Actually, 
Kramers considerations strictly adhered to {\it classical motion}, in
which case the diffusion coefficient is given by the Einstein relation
$D_{pp}=\gamma T$; in the quantum case another term appears.
As is well known, Kramers has used this equation to calculate the decay
rate for a meta-stable situation like fission, in generalization of the
famous Bohr-Wheeler formula. In these days the origin of dissipation
was attributed to the strong "correlations" among the nucleons, as they
can be understood within or follow from N. Bohr's compound
nucleus---and which by definition occur "incoherently".

\ref{hofrep}
\edef\nhofrep{\the\refnumber}
\ref{bloskalswia}
\edef\nbloskalswia{\the\refnumber}
\ref{vankampen}
\edef\nvankampen{\the\refnumber}
In this lecture we want to look at this transition from "independent
particle motion to collisional dominance" in the view of the "linear
response approach", a complete version of which can be found in
[\nhofrep]. This discussion will be complemented by presenting new
aspects in the relation to wall friction, following the more recent
considerations of the group of W.J. Swiatecki, J. Blocki and others
(see [\nbloskalswia]). The applicability of linear response
theory may be understood by the following arguments. First, one
my note that the solution of (\ntraneq) can be written in the
following way
\equorder{propagdens}
\propagdens
\edef\npropagdens{\the\equnumbera}
where $K(Q,P,t;Q_0,P_0,t_0)$ is interpreted as the conditional
probability for the system to move from $Q_0,P_0$ at $t_0$ to 
$Q,P$ at time $t$. On both sides of this relation the distribution $f$,
the "joint probability", may be replaced by conditional probabilities
defining the transition say from a $t_0$ to the final time $t$
through an intermediate step at $t_1$. The resulting relation is
nothing else but the Chapman-Kolmogorov equation.  (For a
discussion of such general properties we may refer to the book by van
Kampen [\nvankampen]).  The procedure just described may be
repeated as often as one likes.  Starting from the {\it given}
equation (\ntraneq) one may introduce {\it arbitrarily small}
time steps $\delta t$ in completely rigorous manner. The reason
for this behavior is found in the fact that this equation
(assumedly) describes a genuinely Markovian process.

It is exactly at this stage where a possible justification of a linear
response approach has to set in. In essence this requires two steps.
First of all, if the $\delta t$ may be chosen to be {\it sufficiently
small on the collective time scale} one may construct the
$K(Q,P,t;Q_0,P_0,t_0)$ by describing collective motion {\it locally to
harmonic order}. Secondly, if the $\delta t$ is {\it large enough on
the microscopic scale} the dynamics of the intrinsic degrees of freedom
does not have to be followed in complete detail. Using such hypothesis
it is possible to construct the form of $K(Q,P,t;Q_0,P_0,t_0)$ 
explicitly and to derive microscopic expressions for the individual
transport coefficients. Moreover, this procedure even allows one to
generalize Kramers' equation to include quantum effects, which show up,
first of all, in generalized diffusion coefficients; for details see
[\nhofrep]. One step necessary in this direction is to interpret the 
$K(Q,P,t;Q_0,P_0,t_0)$ from above as a "propagator" for Wigner
functions.

It is clear that this {\it locally harmonic approximation (LHA)} is closely
related to properties which one expects to hold true for Fokker-Planck
equations. Nevertheless, there are various ways to check that the goal
set at the beginning is actually reached in the very end. For instance,
it is possible to see a) whether the local propagators observe Markovian
behavior,  or b) whether or not the process is indeed "diffusive". As
it turns out, the latter feature ceases to be given for {\it unstable}
modes at low temperatures. We will not have time to go any further into
these questions. We shall, however, be able to touch upon another
condition for the LHA to be valid, the "smoothness" of the transport
coefficients as function of the collective variables.

\bigskip\goodbreak
\leftline{{\bf 2 Linear response theory for collective motion}
\hfill }
\medskip

\ref{kidhofiva}
\edef\nkidhofiva{\the\refnumber}
In the sequel let's suppose to be given a Hamiltonian
$\ham{Q}$ for the nucleons' dynamics in a deformed mean field, with the
deformation being parameterized by the shape variable $Q$, whose
average $\bra \ham{Q} \ket$ represents the total energy of the system
$E_{\rm tot}$ (eventually {\it including} both the Strutinsky
re-normalization as well as "heat").  The equation for average
motion (EOM) for $Q(t)$ can then be constructed from energy
conservation. From Ehrenfest`s equation it follows:
\equorder{eqofmotim}
\eqofmotim
\edef\neqofmotim{\the\equnumbera}
All one needs to do to get the equation of motion for $Q(t)$ is to
express the average $ \left< \ffield{Q} \right>_{t}$ as a functional of
$Q(t)$. Following the scheme of the LHA one may
expand the $\hat H(Q)$ around any given $Q_0$ to have:
\equorder{hamilapp}
\hamilapp
\edef\nhamilapp{\the\equnumbera} 
The effects of the coupling term
$(Q(t)-Q_0)\h{F}$ may now be treated by linear response theory,
exploiting as a powerful tool the causal response function $\rft$ 
\equorder{defrest}
\defrest
\edef\ndefrest{\the\equnumbera}
Here, the time evolution in  $\hat F^{I}(t)$ as well as
the density operator $\equiop$ are determined by $H(Q_{0})$. The
$\equiop$ is meant to represent thermal equilibrium at $Q_0$ with
excitation being parameterized by temperature or by entropy.
After some lengthy derivation one sees that the frequencies for local
motion are given by the secular equation $  \chi(\om)+k^{-1}=0$,
which actually determines the poles of the {\it collective} response
$  \respc = {\chi(\om) /( 1+k\chi(\om))}$. Different to common
approaches but most important, in our case the coupling constant $k$
appearing here is a {\it derived} quantity, given in the end by
$-k^{-1} = \left<{\partial^2\hat H/ \partial Q^2}
    \right>^{\rm qs}_{Q_0,T_0} + \chi(0)-\chi^{\rm ad}   =
   \left.{\partial^{2}E(Q,S_{0})/ \partial  Q^{2}}\right\vert_{Q_{0}}
   +\chi (0)$,
with $\chi(0)$ being the static response, $\chi^{\rm ad}$ the
adiabatic susceptibility and $E(Q,S_{0})$ the quasi-static
energy at given Q and fixed entropy $S_0$. Finally, the
transport coefficients for average motion can be introduced
whenever it is possible to approximate the $\respc$ by an
oscillator response function $\chiosc$, in the sense of having
\equorder{tranosc}
\tranosc
\edef\ntranosc{\the\equnumbera}
The $ \fext(\om)$ represents an "external" field which couples to our
system through a term $ \fext(t)\hat F$. Self-sustained motion corresponds
to $ \fext(t)=0$, in which case the {\it total} energy must be conserved
(according to (\neqofmotim)). As shown first in [\nkidhofiva] (for the 
damped self-consistent case) the equation $ {d E_{\rm tot}/dt}  =0$
can be rewritten as
\equorder{decolldiff}
\decolldiff
\edef\ndecolldiff{\the\equnumbera}
which correctly expresses the exchange between collective motion into
heat. (The $\om_{1}$ represents one of the possible (complex!)
frequencies of the system, as determined from the secular equation).

\medskip\goodbreak
\leftline{{\bf 3 Forced energy transfer to a system of independent
particles}
\hfill }
\medskip
As is clearly seen from (\ndecolldiff), the friction force
parameterizes that energy which is transfered {\it irreversibly} 
to the intrinsic system. Let us study this feature within a simple
model, with the simplifications consisting first of all in neglecting
self-consistency. This means that we take a nucleus at given
deformation $Q_0$ which is {\it exposed to a time dependent external
field} at some fixed polarization. We may thus use a Hamiltonian of
the type given in (\nhamilapp), where the $\hat F$ is chosen to
represent this polarization, but {\it where the last term on the right
is neglected}. The $Q(t)-Q_0=q(t)$ is then a truly {\it external} quantity,
which shall be called $q(t)$ in the sequel, and which is {\it not
subject} to a subsidiary condition of the type $k \fmb_t=q(t)$,
which follows from (\neqofmotim) and (\nhamilapp). As another
important simplification we will assume the $\hat H(Q_0)$ to represent
the ensemble of {\it independent particles} as given by the deformed
shell model at {\it zero temperature}.

Such a system has been studied in a series of papers which aimed at a
new understanding of the physics of wall friction (see [\nbloskalswia]
and references given there). The time dependence of the $q(t)$ was
assumed to be of the form  $q(t)=q_0 \sin(\Omega t) $ and the system
was followed for one period simulating the solutions of the
Schr\"odiger equation numerically.

Let us examine this problem within linear response theory. The energy
transfered to the intrinsic degrees of freedom within one cycle
 may be evaluated from the following well known formula:
\equorder{avernlos}
\avernlos
\edef\navernlos{\the\equnumbera}
The last expression is correct only for the truly periodic field. To
derive it one needs to use the fact that the Fourier transform of the response
function $\tilde \chi (t)$ may conveniently be split into real and
imaginary parts, $\chi(\omega)=\chi^\prime(\omega) + i
\chi^{\prime\prime}(\omega)$, where (for real $\om$)
$\chi^\prime(\omega)$ is an even function and
$\chi^{\prime\prime}(\omega)$ an odd one. For this reason only
the latter one survives after integrating twice over time.  (For more
details on these features see e.g.[\nhofrep]).

This result may be compared with those of [\nbloskalswia],
we simply need to identify $\overline{\Delta E_{int}}$ with their
$\Delta E$  and calculate the total unperturbed energy $E_0$ as the sum
over single particle energies. However, one may as well go ahead
and introduce already here a friction coefficient by the
following reasoning. As the $\overline{\Delta E_{int}}$ measures
the change of energy during one period of vibrations, one may
simply divide by the length $T=2\pi/\Omega$ of that period. In this way one
gets:
\equorder{avenlosfric}
\avenlosfric
\edef\navenlosfric{\the\equnumbera}
with $v \equiv \dot q$, $\dot q_0=\Omega q_0$ and ${\overline
v}^2=(\dot q_0)^2/2$. We have identified the friction coefficient as
\equorder{fricenlo}
\fricenlo
\edef\nfricenlo{\the\equnumbera}
with the function $\Phi^{\prime\prime}(\Omega)$ being the so called
relaxation function. Notice that the frequency appearing here is the
(real) one given by the external field. This is very different from the
form indicated in (\ndecolldiff). As mentioned there, the $\om_{1}$ is
the actual, complex frequency the collective motion has around the
$Q_0$.  Incidentally, a form of the type ${\chi^{\prime\prime}(\Omega)
/ \Omega}$ may appear (for friction) even within the linear response
formulation as described before, but only if the coupling between
collective and intrinsic motion is treated perturbatively, see section
3.3.2 of [\nhofrep]. In case of small frequencies one may apply the so
called zero frequency limit $\gamma(\Omega=0)$. 

In Figs.1 and 2 we present numerical results for the quantity 
$\Delta E /E_0 = \overline{\Delta E_{int}} /E_0 $ for the case
of quadrupole excitations. They were calculated on
the basis of our formula (\navenlosfric) but for the same system 
as in [\nbloskalswia], namely independent particles in a Woods-Saxon
potential (of an un-physically large depth to decrease the escape
probability). All parameters are chosen like there, which means
that the $\eta$ can approximately be written as $\eta\approx
0.02269 \hbar \om$.  
As the most striking difference to the (quantal) results presented in 
Fig.1 of [\nbloskalswia], in our case we observe strong oscillations with
$\eta$, which represent nothing else but the typical strength function
behavior. (These functions are smooth in omega simply because we
averaged the delta functions over an interval of 0.1 MeV).
In both figures we show as the straight line marked
with dots the result one gets in case that this energy transfer
is calculated with wall friction. Apparently the latter result
can be obtained at best after performing some averages.

\ref{hoivyanp}
\edef\nhoivyanp{\the\refnumber}

Indeed, it has been shown in [\nhoivyanp] that the friction
coefficient obtained within linear response theory (in the zero
frequency limit) becomes close to the one of the wall formula
after applying smoothing procedures in the sense of the
Strutinsky method. This features goes along very nicely with the
claim that wall friction represents the "macroscopic limit", for
a system of independent particles (for an extensive discussion
of this topic see [\nhofrep]). The same feature is seen here
for the $\gamma(\Omega)$ of (\nfricenlo). In Figs.1 and 2 we
present curves obtained from applying Strutinsky smoothing to
the microscopic evaluations: For the dotted lines the averaging
interval was 5 MeV, for the short dashed ones 10 MeV and for the
long dashed ones 20 MeV. From Fig.1 it is seen that and how 
smoothing leads to results similar to that of the wall formula.
This calculation corresponds to the case where $Q_0$ stands for
a spherical deformation, the same situation which has been
considered also in [\nhoivyanp]. The one presented in Fig.2
corresponds to a case where the unperturbed system has a sizable
octupole deformation of $\alpha=0.3$. In this case the wall
formula is not recovered (at least not the one corresponding to
the spherical configuration used in the figure, whereas there
could be some small dependence on deformation).  We may say that
similar results are obtained for vibrations of other
multi-polarity. As an interesting feature we may note that
"macroscopic" friction is smaller the more complex the
microscopic strength distribution is. We would not like to
speculate whether or not this fact is related to an increase of
chaotic behavior of the nucleonic degrees of freedom.

In [\nbloskalswia] only forced vibrations around the sphere were
considered. As can be seen from their Fig.1, the simulations of
the Schr\"odinger equation indicate a straight behavior of the
functional dependence of $\Delta E/E_0$ on $\eta$ somewhat below
the wall formula. In such a non-perturbative calculation an
average over a full cycle will differ in at least one respect
from our procedure. Starting from formula (\neqofmotim) (which is
a correct one), the $\left< \ffield{Q} \right>_{t}$ will
{\it no longer be a linear} functional of $q(s)$. If one still
were expressing this quantity by an integral like $\int_{-\infty}^{\infty} 
   ds\, \tilde \chi(t-s)q(s)$ the $\tilde \chi(t-s)$ itself
would have to be a complicated functional of  $q(s)$. Apparently, it
behaves such that the average over the amplitude of oscillation
(in the deformation degree of freedom) in the end leads to a
linear relation with $\eta$. It seems to us that this average
may in a sense be considered analogous to our averaging in the
spectrum, as is done in the Strutinksy method. As a matter of fact, experience tells one that averaging in
energy $e$ over an interval of $\gamma^{av}_e=10$ MeV corresponds
to an average in Q over a $\gamma^{av}_Q= \gamma^{av}_e / e_F
\approx  1/4$ --- which corresponds nicely to
the amplitude chosen in [\nbloskalswia]! 

\medskip\goodbreak
\leftline{{\bf 4 The influence of collisional damping on transport
properties} 
\hfill}
\medskip
\ref{bertschrel}
\edef\nbertschrel{\the\refnumber}
From the discussion of the last section it is clearly seen that for
nuclear collective motion it is {\it not possible to justify a local
friction force within the mere picture of independent particles}.
For such a model one has to employ averaging procedures of one
kind or other. Moreover, we have observed that quite large intervals 
in the averaging parameters are involved if for the latter one chooses
energy. This fact clearly hints to an inherent deficiency of the
underlying model: At the excitations which are at stake in common
experimental situations the picture of particles moving in a mean field 
{\it without "collisions" does not apply}! For this reason the notion of
the  $\hat H(Q_0)$ to be simply given by the deformed shell model has
been given up a long time ago whenever transport properties where
calculated within the linear response approach (see [\nhofrep] for a
detailed discussion). Instead it was assumed that the particles are
dressed by self-energies having both real and imaginary parts:
$ \self(\om\pm i \epsilon,T) = \selfre(\om,T) \mp {i\over 2} \Gamma(\om,T) $.
The intrinsic response functions are then calculated after replacing 
the single particle strength  $\varrho_k(\om)=
2\pi\;\delta(\hbar \om - e_k) $ by
\equorder{dspspecdensgam}
\dspspecdensgam
\edef\ndspspecdensgam{\the\equnumbera}
with the $\mu$ being the chemical potential.
 
The $1/\Gamma_0$ represents the strength of the "collisions", viz of
the coupling to more complicated states. The cut-off parameter $c$
allows one to account for the fact that the imaginary part of the
self-energy does not increase indefinitely when the excitations get
away from the Fermi surface. Both parameters are not known precisely,
but from experience with the optical potential and the effective masses
the following range of values can be given: $0.03 MeV ^{-1} \le
\Gamma_0^{-1} \le  0.06 \quad MeV ^{-1}$ and $ 15 MeV  \le c \le  30
MeV$.  Neglecting the $\om$ dependence of $\Gamma$ and putting $c \to
\infty$ these values lead to a average relaxation
time for single particle motion $\tau_{\rm int} = \hbar/\Gamma$ which is in
accord with the estimate given in [\nbertschrel].

\ref{ivhopaya}
\edef\nivhopaya{\the\refnumber}

In Fig.3 we present  calculated along a
fission path of $^{224}{\rm Th}$ for different temperatures
(whose values are given in MeV). All curves except the one
marked by triangles are identical to those of  Fig.13 of
[\nivhopaya], where for the deformed shell model the
Pashkevich code has been employed; details can be found in
[\nivhopaya]. It is seen (i) that this ratio $\gamma /M$ does not
change very much with the collective variables as soon as $T$ is
of the order of 2 or larger, and (ii) that it {\it increases
with T} (for reasons given below, the "heat pole" contribution has
been removed in this calculation). For larger $T$ the ratio is
of the order as predicted by the wall formula (for $\gamma$)
plus the one of irrotational flow for the inertia. The reason
for this behavior is due to the fact that with increasing $T$
the residual interactions become more and more important, with the two
implications of a) smoothing out details of shell structure and
b) making the microscopic mechanism of dissipation more
effective. 

\ref{ivhopair}
\edef\nivhopair{\the\refnumber}

For $T=1$ we have included a still preliminary result of an
extension of our theory to the inclusion of pairing
correlations. As expected the latter reduce the influence of
shell effects, albeit details still will have to be clarified
further [\nivhopair]

\medskip\goodbreak
\leftline{{\bf 5 The role of symmetries and the heat pole for
nuclear friction} 
\hfill}
\medskip
Above it has been indicated that for the calculations presented
in Fig.3 a particular contribution to friction was discarded.
This shows up at finite $T$ and is related to an
interesting quasi-static property which in turns is dominated by
the influence of symmetries (for a detailed discussion see
[\nhoivyanp, \nhofrep]). Let us demonstrate these features
with the help of the zero frequency limit of friction.
To sufficient accuracy the latter can be written as
\equorder{zefridisflu}
\zefridisflu
\edef\nzefridisflu{\the\equnumbera}
On the very right the correlation function has been introduced which is
related to the response function by the famous fluctuation dissipation
theorem (FDT) $\hbar \drf(\om)=
     \tanh\left({\hbo/ 2T}\right) \psi^{\dpr}(\om)$. The definition of 
$\psi$ is similar to that given for $\chi$ in (\ndefrest), with two
important exceptions: The commutator is to be replaced by an {\it
anti-commutator} and from the operator $\hat F$ one has to subtract its
unperturbed average value $\fmb$. The general microscopic
expression for $\psi^{\dpr}(\om)$ is
\equorder{corrsingreg}
\corrsingreg
\edef\ncorrsingreg{\the\equnumbera}
with the $\;_R\psi^{\dpr }(\om)$ being regular at $\om=0$. The
$\chi^{\rm T}$ is the {\it isothermal susceptibility} which measures
how the (quasi-)static expectation value $\fmb^{qs}$ changes with $Q$
if the temperature is kept constant. The singularity at $\om=0$ is
called "heat pole", in analogy to a similar pole in the density
density strength distribution for infinite matter being responsible for
heat diffusion there. A structure like that given in (\ncorrsingreg) is
obtained only in the strict case of pure Hamiltonian dynamics (i.e.
when the correlation function is formally calculated in the basis of
exact eigen states). Within our approximation of collisional damping
the heat pole changes like
\equorder{heatplor}
\heatplor
\edef\nheatplor{\the\equnumbera}
Both $\psi^0$ as well as $\Gamma_T$
have been calculated numerically in [\nivhopaya]. The result 
for $\Gamma_T$ follows closely the following simple rule $\Gamma_T\approx
2\Gamma(\mu,T) \approx 2 T$, which is valid over the very large
range of temperatures of up to about $10$ MeV.

\ref{aynoediabfri}
\edef\naynoediabfri{\the\refnumber}

When applied to (\nzefridisflu) one sees that the heat pole implies the
following contribution to friction
\equorder{zerfriheat}
\zerfriheat
\edef\nzerfriheat{\the\equnumbera}
Estimating  $\chi^{\rm T} - \chi(0)$ simply
in the independent particle model this component of friction turns into
the one found first by Ayik and N\"orenberg within the model of DDD
[\naynoediabfri]. In [\nhoivyanp] this
form has been evaluated as function of temperature for all $T$. 
A slightly modified version is shown in Fig.4. The fully drawn
line and the dashed one correspond the ${_0\zerofri}$ of
(\nzerfriheat) with the $c$ of (\ndspspecdensgam) put equal to
$c=20$ MeV and $c\to \infty$, respectively. The curve with the
heavy dots corresponds to the contribution of the remaining part
of the correlation function.  As demonstrated in [\nhoivyanp]
(see also [\nhofrep]), the distinction of the two contributions
can simply be made in terms of the matrix elements of the
(one-body) operator $\hat F$ with the shell model states. The
${_0\zerofri}$ is solely to be associated to the {\it diagonal}
elements. That they may lead to dissipation, nevertheless, (and
thus to entropy production) is due to the effects of
"collisions".

From Fig.4 it is seen that at smaller $T$ the ${_0\zerofri}$
takes on very large values. They actually exceed several times
that of wall friction (shown by the horizontal line), and they
seem to be too large as required by experimental evidence (see
[\nhofrep]). The reason can be traced back to the following
properties of static susceptibilities. Let us first rewrite the
difference appearing in $\psi^0$ as given by the left part of
the following equation:
\equorder{sepchiadtis} 
\sepchiadtis
\edef\nsepchiadtis{\the\equnumbera}
The difference of the adiabatic to the isothermal susceptibility
can be seen to be small in the nuclear case; it is proportional
to the square of the cross derivative of the free energy with
respect to $Q$ and $T$, a quantity which for the system
underlying Fig.4 even vanishes.  So the culprit must be the
$\chi^{\rm ad} - \chi(0)$! However, this difference is known to
vanish for truly ergodic systems, namely systems whose states
are non-degenerate. As an additional condition one only needs to
have a sufficiently narrow distribution of the energies of the
excited states.

Apparently, these conditions are {\it not met} for the case
shown in Fig.4, and most likely {\it both} of them are violated.
First of all, the microscopic evaluation of the matrix elements
is dominated by the model of independent particles, with
all the many degeneracies appearing there. Secondly,
applying the canonical distribution to parameterize the thermal
excitations of a nuclear system the spread in energy is
exaggerated artificially. On the other hand, there is little
doubt that the true compound configurations will remove these
spurious contributions. First of all, because a consideration of
the compound states will require a more correct treatment of the
residual interactions. In this way the many artificial
degeneracies of the deformed shell model will be removed.
Secondly, thermal excitations will have to be treated on the basis
of the micro canonical ensemble. To simply simulate these
effects one may just apply the reduction indicated in
(\nsepchiadtis) to remove the unphysical contribution from the
heat pole.

\ref{yaivho}
\edef\nyaivho{\the\refnumber}
\medskip\goodbreak
\leftline{{\bf 6 Dissipation within Landau theory}
\hfill}
\medskip
As seen above, the friction coefficient tends to decrease with
$T$ at larger temperature. This feature is evident for the component
${_0\zerofri}$ (see (\nzerfriheat) and Fig.4), but as discussed
in [\nhoivyanp] it will eventually hold true also for the other
component (see also [\nyaivho]) under certain circumstances
(like approximating the imaginary part of the self-energy in
"common" relaxation time approximation (with $c=\infty$)).

\ref{mahokosh}
\edef\nmahokosh{\the\refnumber}
\ref{maghof}
\edef\nmaghof{\the\refnumber}

Such a behavior with $T$ reminds one of the two body viscosity
of hydrodynamics. In [\nmahokosh] a model has been suggested in
which the intrinsic dynamics is described by the Landau-Vlasov
equation and where the finiteness of the system exhibiting shape
dynamics is introduced through special boundary conditions. In
Fig.5 we present a calculation of the friction coefficient (as
function of $T$) for quadrupole vibrations about a sphere, done
within an extension of this model. The dashed and the fully
drawn lines correspond to the hydrodynamical limit, for two
different choices of the parameter $c$ entering the relaxation
time used in the collision term of the Landau-Vlasov equation.
The squares correspond to contributions from different peaks in
the correlation function, where the full ones are suposed to
correspond to the analog of the "heat pole". Similarities with
the behavior shown in Fig.4 are evident. So far however, it is
yet unclear exactly where the contribution comes from, solely
from the difference $\chi^{\rm T} - \chi^{\rm ad}$ as it should,
or whether also in this model there is the spurious effect
coming from a non-vanishing difference $\chi^{\rm ad} -
\chi(0)$. Further studies are under way.


\medskip\goodbreak
\leftline{{\bf 7 Summary}
\hfill}
\medskip
To describe collective motion as a Markovian transport process
one needs to be able to define transport coefficients which vary
smoothly with the macroscopic variables, which by the way have
to include the parameter which measures the thermal excitation.
We have demonstrated that such a condition can hardly be
fulfilled within the picture of the deformed shell model. On the
other hand, we have shown that residual interactions may do
the job, the better the larger the thermal excitation. At
present the situation is less clear at smaller temperatures. 
Whether or not pairing alone will do is currently under
investigation. It may well be, however, that even in this regime
one may want to include more of the configurations as given by
the nuclear compound model.

\vfill
\eject
\bigskip\goodbreak
\leftline{{\bf References}}
\medskip
\item{1)} H.A. Kramers, Physica 7 (1940) 284
\item{2)} H. Hofmann, Phys. Rep. 284 (4\&5), (1997) 137-380
\item{3)} J. Blocki, J. Skalski and W.J. Swiatecki, 
    Nucl. Phys. A594 (1995) 137
\item{4)} N.G. van Kampen: "Stochastic processes in physics and
      chemistry ", North-Holland, 1981, Amsterdam
\item{5)} D. Kiderlen, H. Hofmann and F.A. Ivanyuk,  
     Nucl.Phys. A550 (1992) 473
\item{6)} H. Hofmann, F.A. Ivanyuk and S. Yamaji, Nucl. Phys. 
    A 598 (1996) 187
\item{7)} G.F. Bertsch, Z.Phys. A289 (1978) 103
\item{8)} F.A. Ivanyuk, H. Hofmann,  V.V. Pashkevich and S.
    Yamaji, Phys. Rev. C 55 (1997) 1730
\item{9)} F.A. Ivanyuk and H. Hofmann, to be published
\item{10)} S. Ayik and W. N\"orenberg, Z. Phys. A309 (1982) 121
\item{11)} S. Yamaji, F.A. Ivanyuk and H. Hofmann, 
     Nucl. Phys. A612 (1997) 1
\item{12)} A.G. Magner, H. Hofmann, V.M. Kolomietz and S. Shlomo,
    Phys.Rev.C 51 (1995) 2457
\item{13)} A.G. Magner and H. Hofmann, to be published

\vfill
\eject
\medskip
\leftline{{\bf Figure captions}}
\medskip
\parindent = 0 pt
Fig.1: Average energy transfer to a spherical system of independent
particles by an external quadrupole perturbation, calculated
within linear response. Otherwise the same picture is adopted as
in [\nbloskalswia].
\bigskip
Fig.2: Same as for Fig.2 but for a system with octupole deformation.
\bigskip
Fig.3: Ratio of friction to inertia along the fission path of 
$^{224}{\rm Th}$ (for details see text).
\bigskip
Fig.4: The contribution of the "heat pole" to friction for
collective quadrupole oscillations of a system of particles in
a square well potential (see text).
\bigskip
Fig.5: Friction for quadrupole oscillations calculated from a
Landau-Vlasov approach to a finite nucleus.

\vfill
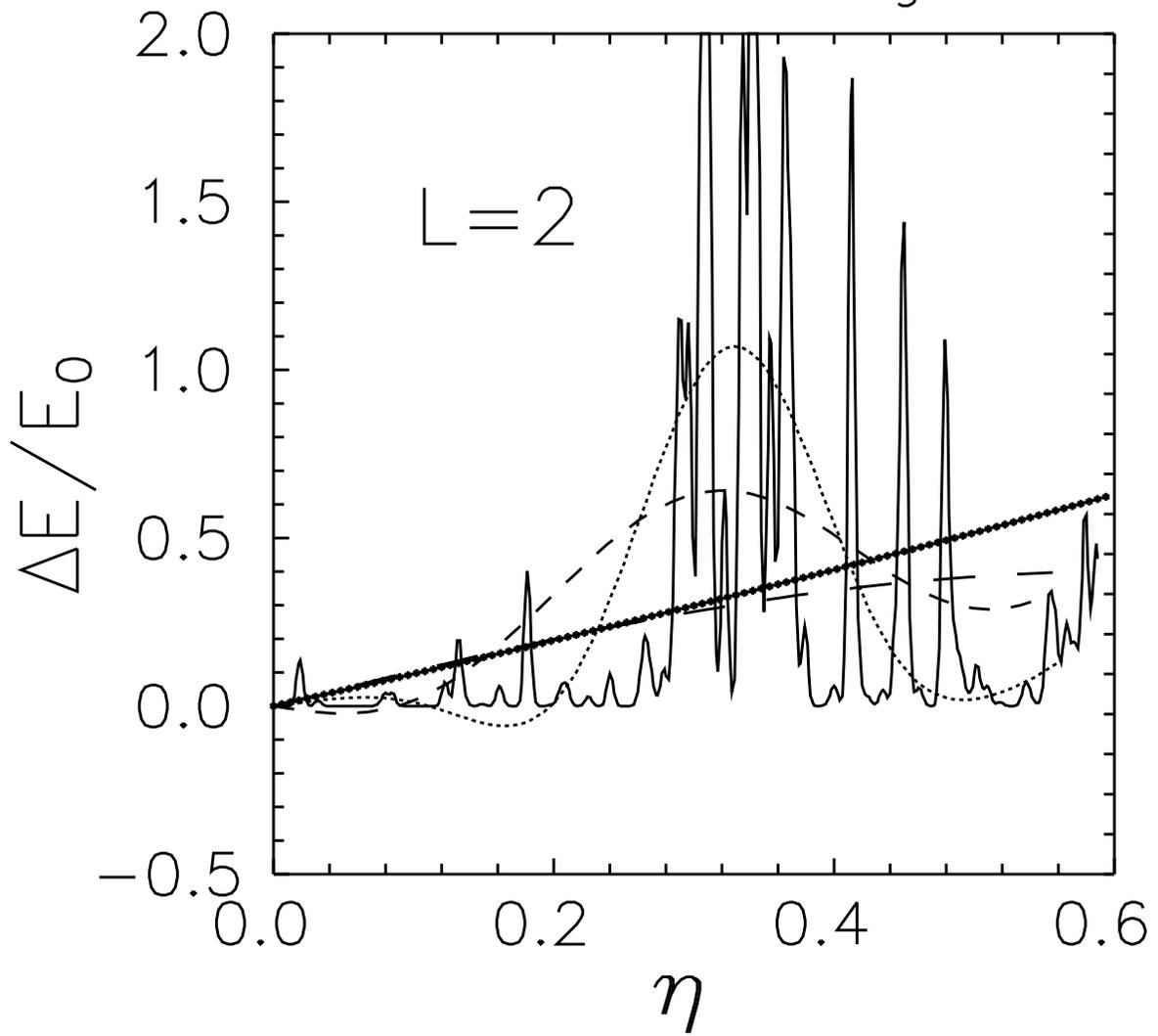
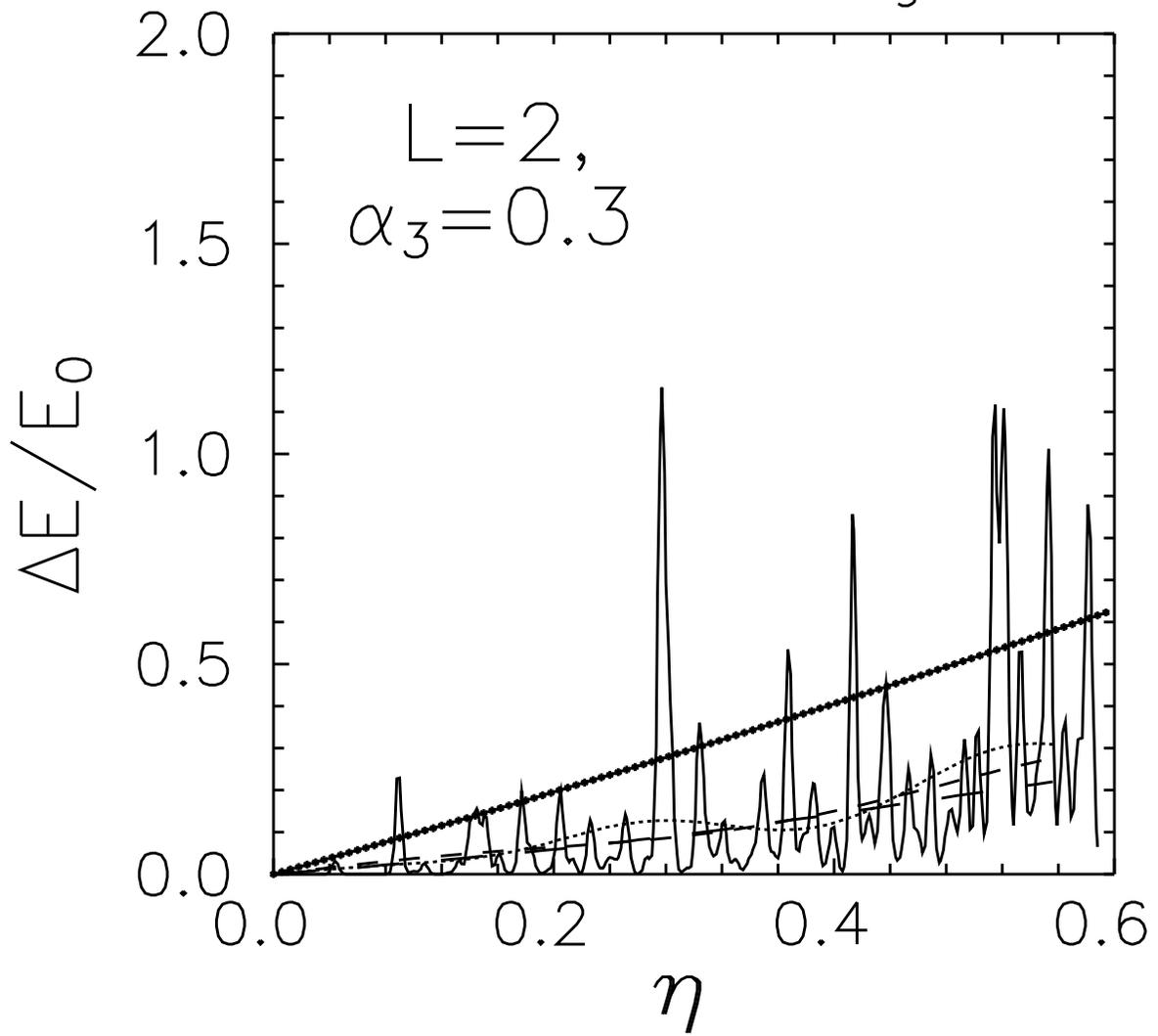
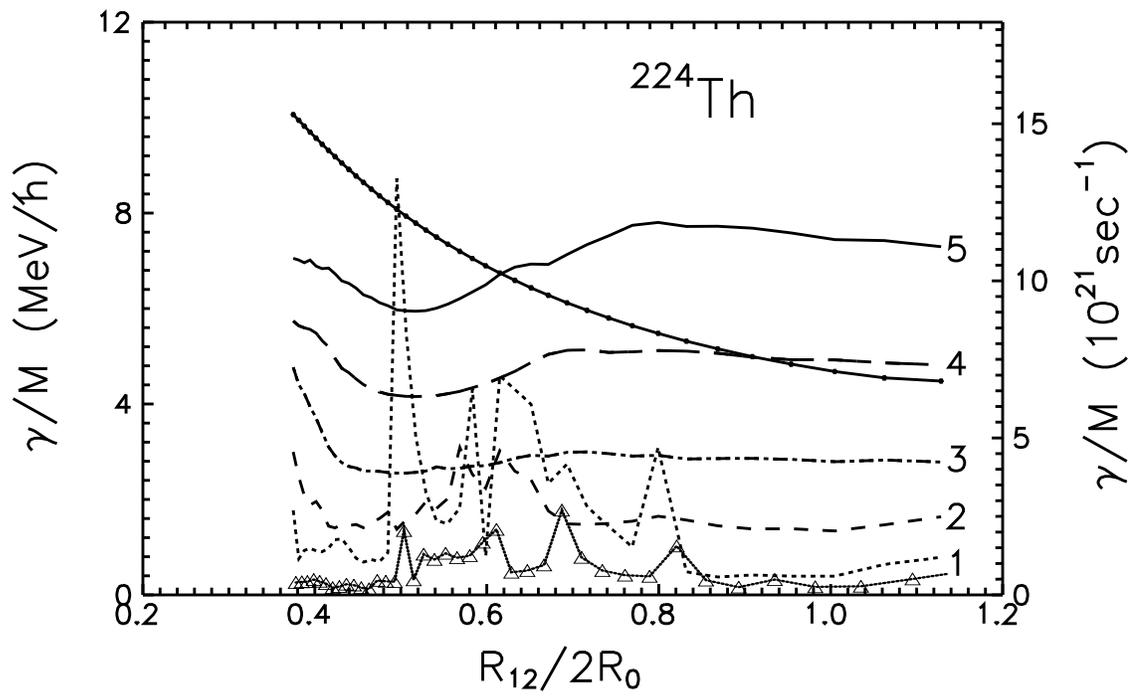
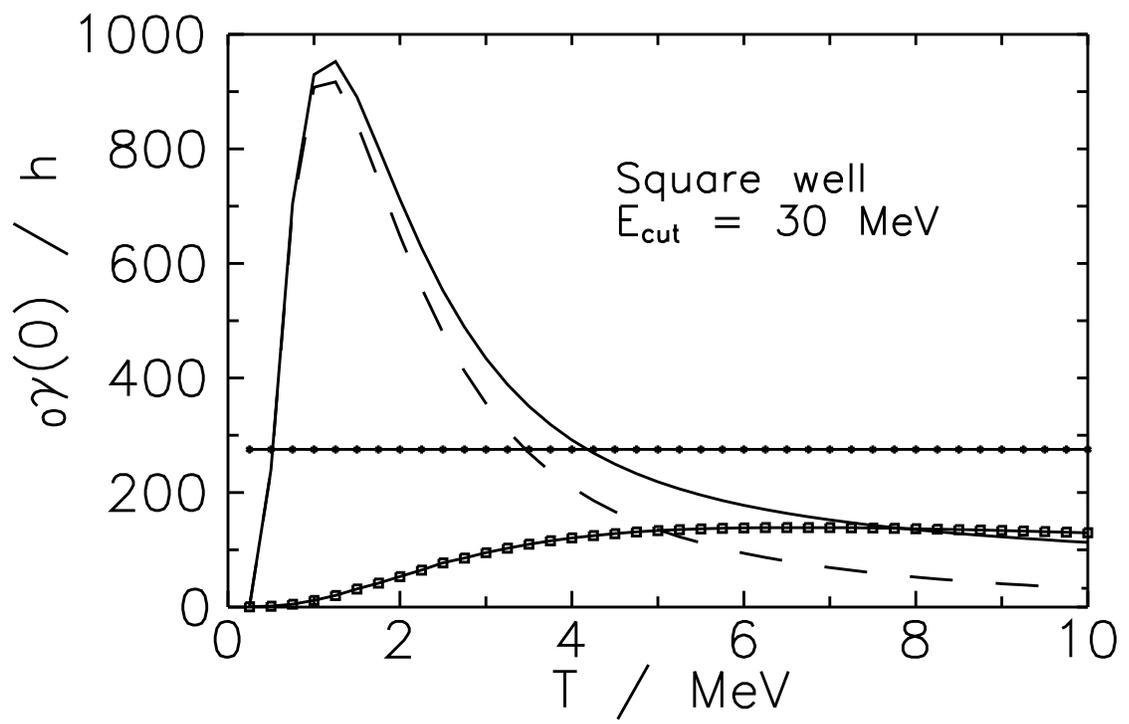
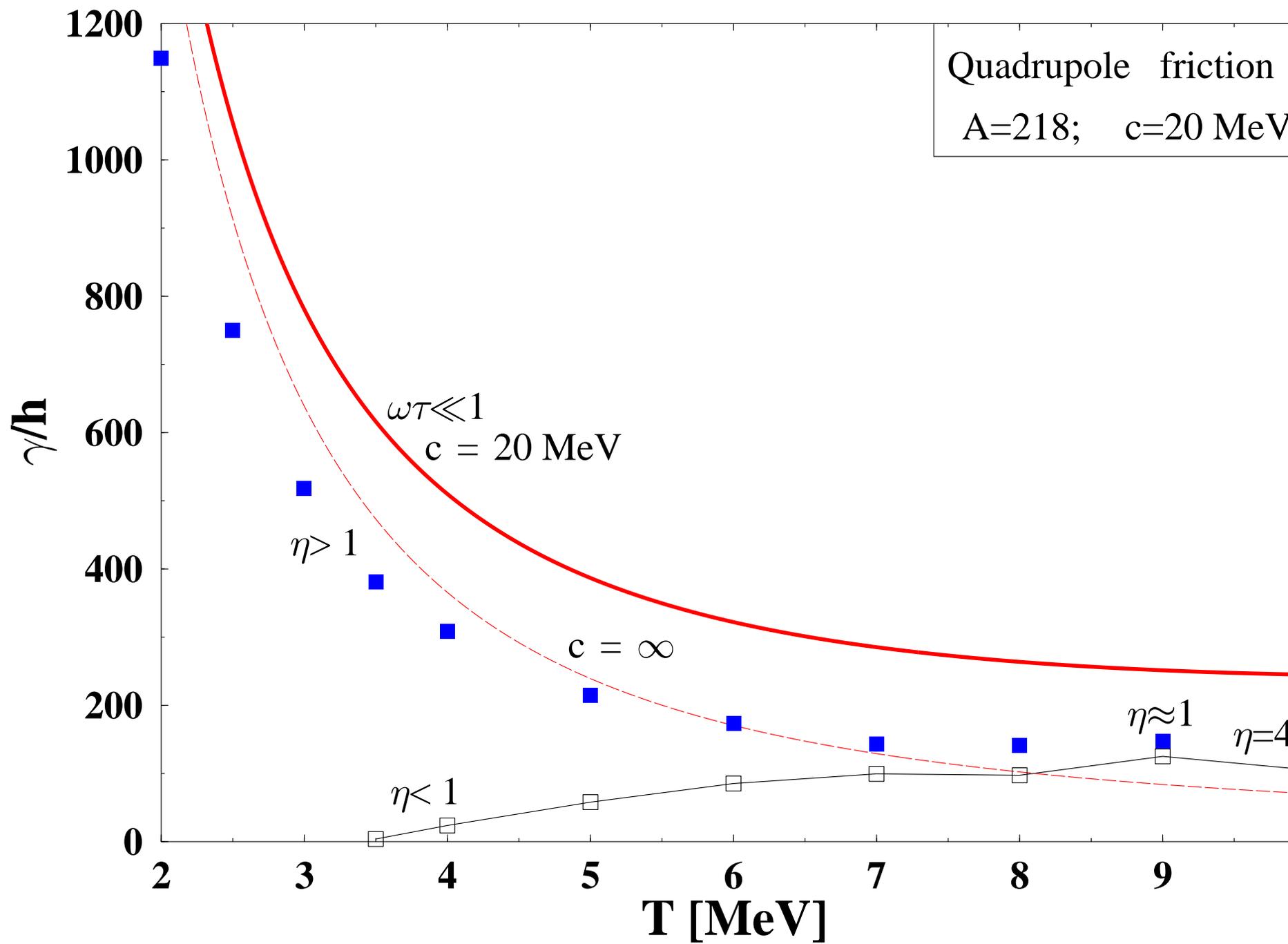
\end